\documentclass[aps,prc,twocolumn,nofootinbib,preprintnumbers,amsmath,amssymb]{revtex4}

\usepackage{graphicx}
\usepackage{dcolumn}
\usepackage{bm}
\usepackage{epsfig}
\usepackage{longtable}

\def\fun#1#2{\lower3.6pt\vbox{\baselineskip0pt\lineskip.9pt
\ialign{$\mathsurround=0pt#1\hfil##\hfil$\crcr#2\crcr\sim\crcr}}}

\begin{document}

\preprint{}

\title{
Precise measurement of $\alpha_K$ and $\alpha_T$ for the 39.8-keV $E$3 transition in $^{103}$Rh: Test of internal-conversion theory
}

\author{N. Nica}
\email{nica@comp.tamu.edu}

\author{J.C. Hardy}
\email{hardy@comp.tamu.edu}

\author{V.E. Iacob}

\author{V. Horvat}

\author{H.I. Park}

\author{T.A. Werke}

\author{K.J. Glennon}

\author{C.M. Folden III}

\author{V.I. Sabla}
\altaffiliation {REU summer student from Middlebury College, Middlebury VT 05753}

\author{J.B. Bryant}
\altaffiliation {REU summer student from University of Central Arkansas, Conway AR 72035}

\author{X.K. James}
\altaffiliation {REU summer student from University of Wisconsin, La Crosse WI 54601}

\affiliation{ Cyclotron Institute, Texas A\&M University, College Station, Texas 77843, USA}
\homepage{http://cyclotron.tamu.edu/}

\author{M.B. Trzhaskovskaya}
\affiliation{Petersburg Nuclear Physics Institute, Gatchina 188300, Russia}

\date{\today}

\begin{abstract}
Neutron-activated sources of $^{103}$Ru and $^{103}$Pd both share the isomeric first-excited state in $^{103}$Rh as a daughter product.  From independent
measurements of both decays, we have measured the $K$-shell and total internal conversion coefficients, $\alpha_K$ and $\alpha_T$, for the 39.8-keV
$E$3 transition, which de-excites that state in $^{103}$Rh, to be 141.1(23) and 1428(13), respectively.   When compared with Dirac-Fock calculations, our new
results disagree with the version of the theory that ignores the $K$-shell atomic vacancy, which is consistent with our conclusion drawn from a series of
measurements on high multipolarity transitions in nuclei with higher $Z$.  Calculations that include the atomic vacancy indicate that the transition actually
has a small $M$4 component with mixing ratio $\delta$ = 0.023(5).
 
\end{abstract}

\maketitle

\section{\label{sec:introd} INTRODUCTION}

For over a decade we have been systematically measuring $K$-shell Internal Conversion Coefficients (ICCs) for $E$3 and $M$4 transitions with a precision
of $\pm$2\% or better \cite{Ni04,Ni05,Ni07,Ni08,Ni09,Ni14,Ha14,Ni16,Ni17a,Ni17b}. Our goal throughout has been to test the accuracy of the calculated ICCs \cite{Ba02}
and, in particular, we sought to distinguish between two versions of the theory, one that ignored the atomic vacancy left behind by the emitted electron, 
and another that took the vacancy into account.  We also sought to extend the test over nuclei covering as wide a range of $Z$ values as possible.

Here we present a measurement of the 39.8-keV $E$3 transition in $^{103}$Rh, the ninth in the series and the lowest $Z$ yet.  Before we began the series, there were
very few $\alpha_K$ values known to high precision, so the treatment of the vacancy and the consequent accuracy of the calculated ICCs were controversial
topics \cite{Ra02}.  Today, with our new result there are now twelve $\alpha_K$ values for $E$3 and $M$4 transitions known to better than $\pm$2\%, all but
three being from our work.  They cover the range $45 \leq Z \leq 78$ and, so far, they strongly support the ICC model that includes provision for the atomic
vacancy.

What makes such precise measurements possible for us is our having an HPGe detector whose relative efficiency is known to $\pm$0.15\% ($\pm$0.20\% absolute)
over a wide range of energies: See, for example, Ref.\,\cite{He03}.  By detecting the $K$ x rays and the $\gamma$ ray from a transition of interest in
the same well-calibrated detector at the same time, we can avoid many sources of error in determining $\alpha_K$. Fortuitously, the present measurement also
offers the possibility to determine the total ICC, $\alpha_T$, to the same precision.

Like many experimental ICCs, the $\alpha_K$ and $\alpha_T$ coefficients for the 39.8-keV $E$3 transition in $^{103}$Rh have been measured several
times \cite{Le69,Pe70,Cz75,Ma76,Va79,Sa99} but only once in the past 40 years.  Of the six $\alpha_K$ results, four have uncertainties above $\pm$10\% and
are unable to comment meaningfully on the theory; the fifth and sixth  claim to be $\pm$5\% but one of them \cite{Sa99} quotes a value that agrees
with neither version of the ICC theory, and the other \cite{Cz75}, a value that agrees with the version of the ICC theory that ignores the atomic vacancy, 
which would be striking if true.  There are only two previous $\alpha_T$ results \cite{Cz75,Va79}, one of which agrees
with both versions of the theory, the other with neither.  Furthermore, none of these references acknowledges that the energy of the $\gamma$ ray of interest, 
39.8 keV, is very nearly equal to twice the energy of rhodium $K_{\alpha}$ x rays, $\sim$40.3 keV.  Unresolved random pile-up of x rays could easily have
impacted the $\gamma$-ray peak and distorted the results.  Thus there is good reason to re-measure these ICCs with more modern techniques.

\section {\label{overview} Measurement Overview}

We have described our measurement techniques in detail in previous publications \cite{Ni04,Ni07}
so only a summary will be given here.  If a decay scheme is dominated by a single transition
that can convert in the atomic $K$ shell, and a spectrum of $K$ x rays and $\gamma$ rays is recorded
for its decay, then the $K$-shell internal conversion coefficient for that transition is given by
\begin{equation}
\alpha_K  = \frac{N_K}{N_\gamma} \cdot \frac{\epsilon_\gamma}{\epsilon_K} \cdot \frac{1}{\omega_K},
\label{alpha}
\end{equation}
where $\omega_K$ is the $K$-shell fluorescence yield; $N_K$ and $N_{\gamma}$ are the total numbers of observed
$K$ x rays and $\gamma$ rays, respectively; and $\epsilon_K$ and $\epsilon_\gamma$ are the
corresponding photopeak detection efficiencies.

The fluorescence yield for rhodium has been measured a number of times (see summary in Ref.\,\cite{Ri17}) but with rather modest precision.  However, world
data for fluorescence yields have been evaluated \cite{Sc96} systematically as a function of $Z$ for all elements with $10\leq Z\leq 100$, and $\omega_K$ values
have been recommended for each element in this range.  The recommended value for rhodium, $Z$ = 45, is 0.809(4), which is consistent with the measured values
but has a smaller relative uncertainty. We use this value.

\begin{figure}[t]
\epsfig{file=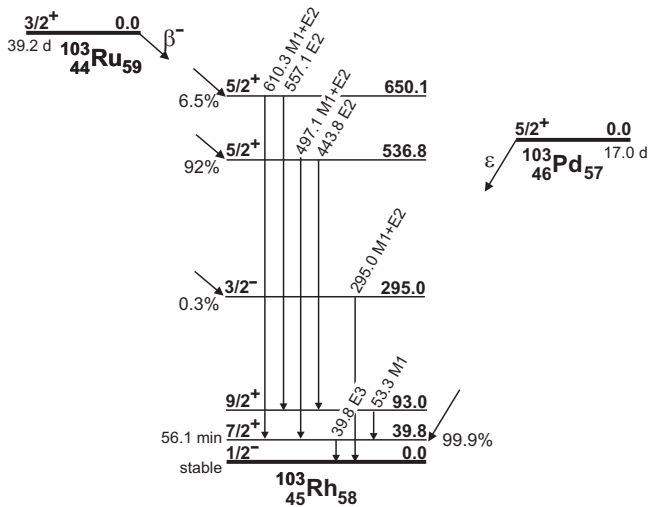,width=8.5cm}
\caption{Simplified decay schemes for the $\beta^-$-decay of $^{103}$Ru and the electron-capture decay of $^{103}$Pd feeding excited states in their common daughter
$^{103}$Rh.  Except for the 39.8-keV transition only electromagnetic transitions with $>$0.1\% $\gamma$-ray intensity are shown.  The data are taken from
Ref.\,\cite{Fr09}, where numerous weaker transitions are recorded.}
\label{fig1}
\end{figure}

Simplified decay schemes are shown in Fig.\,\ref{fig1} for the $\beta^-$ decay of $^{103}$Ru and the electron-capture decay of $^{103}$Pd, which both feed states in
$^{103}$Rh. If the 39.8-keV level is populated by the $^{103}$Ru-decay route, then its decay effectively satisfies the conditions for Eq.\,(\ref{alpha}) since the
295.0-, 443.8-, 497.1-, 557.1- and 610.3-keV transitions have very small $\alpha_K$ values, $\lessapprox$ 0.01; and, although the 53.3-keV transition has a much
larger $\alpha_K$ of 1.81, the transition itself is rather weak.  In total, only about 11\% of the ruthenium x rays do not originate from the 39.8-keV transition,
a small enough amount that it can be reliably determined and subtracted from the measured x-ray intensity before Eq.\,\ref{alpha} is applied, without seriously
degrading the eventual uncertainty on $\alpha_K$.

The three transitions feeding the 39.8-keV level in the decay of $^{103}$Ru also yield a benefit. In the absence of $\beta^-$ feeding to the
39.8-keV state, the total intensity of the electromagnetic transitions populating the state must equal the total intensity of the transition depopulating
it. Consequently we can determine $\alpha_{T39.8}$, the total ICC for the 39.8-keV transition, via the equation,
\begin{equation}
\sum_i(1+\alpha_{Ti}) \cdot \frac{N_{\gamma i}}{\epsilon_{\gamma i}} = (1+\alpha_{T39.8}) \cdot \frac{N_{\gamma 39.8}}{\epsilon_{\gamma 39.8}} ,
\label{alphatot}
\end{equation}
where the sum is over all transitions $i$ that populate the 39.8-keV level. The principal contributors to the sum are the 610.3-, 497.1- and 53.3-keV transitions (see
Fig.\,\ref{fig1}). For the two strongest transitions, their total ICCs, $\alpha_{T497}$ and $\alpha_{T610}$, are calculated to be much smaller than unity, being 0.0053(1)
and 0.0032(1) respectively, independent of whether the atomic vacancy is incorporated or not.  For the 53.3-keV transition, $\alpha_{T53.3}$ is calculated to be larger, 2.08(3)
-- also independent of the treatment of the vacancy -- but the transition is weak enough that its term in the summation contributes only a little more than a percent
to the total. Consequently, from the measured $\gamma$-ray intensities this equation yields a value for $\alpha_{T39.8}$, to which the other conversion coefficients
contribute at the level of a percent or less, and with a negligible effect on the uncertainty.

The situation might appear to be even simpler if the 39.8-keV level is populated uniquely via the $^{103}$Pd-decay route, but it is not.  There is a complication:
In this case, electron capture gives rise to $K$ x-rays in similar numbers to the subsequent internal-conversion process. Fortunately the 39.8-keV level is isomeric, 
so the electron-capture and internal-conversion processes are well separated in time. This means that the $K$ vacancy created by the first process is long filled
before the second takes place. Nevertheless, the contribution from electron capture has considerable impact. 

If we rewrite Eq.\,(\ref{alpha}) to include the contribution from electron capture, we obtain
\begin{equation}
\alpha_K+(1+\alpha_T)P_{ec,K}  = \frac{N_K}{N_\gamma} \cdot \frac{\epsilon_\gamma}{\epsilon_K} \cdot \frac{1}{\omega_K},
\label{alpha2}
\end{equation}
where $P_{ec,K}$ is the probability per parent decay for electron capture out of the atomic $K$ shell; we take its value for this decay to be 0.8595(10) based on
a calculation with the LOGFT code available at the NNDC website \cite{NNDC}.  Although the probability of $K$-capture determines the contribution of $^{103}$Pd
decay to the $K$ x ray peak, it is the electron capture from all shells that determines the population of the 39.8-keV level.  Thus, unlike the $^{103}$Ru decay, 
which yields individual values of $\alpha_K$ and $\alpha_T$ for the 39.8-keV transition, the $^{103}$Pd decay yields a relationship between the two ICCs, as
expressed in Eq.\,(\ref{alpha2}).  This offers a very useful consistency check.

In our experiment, the HPGe detector we used to observe both $\gamma$ rays and $K$ x rays has been meticulously calibrated \cite{Ha02,He03,He04} for efficiency
to sub-percent precision, originally over an energy range from 50 to 3500 keV but more recently extended \cite{Ni14} with $\pm$1\% precision down to 22.6 keV,
the weighted-average energy of silver $K$ x rays.  Over this whole energy region, precise measured data were combined with Monte Carlo calculations from the CYLTRAN code
\cite{Ha92} to yield a very precise and accurate detector efficiency curve.  In our present study, the $\gamma$ ray of interest at 39.8 keV is in the extended
calibration region while the rhodium $K$ x rays, which are between 20 and 23 keV, lie slightly below our existing calibration curve, requiring us to make a very
short extrapolation.  This is easily accomplished with CYLTRAN but leads us to apply a conservative uncertainty of $\pm$1.6\% to the ratio of calculated
efficiencies, $\epsilon_\gamma/\epsilon_K$.

\section{\label{exp} Experiment}

\subsection{\label{sprep} Source Preparation}

We neutron-activated three different sources in the course of this experiment, two of ruthenium and one of palladium.  For all three we used material with
isotopes in natural abundance.  The reactions of interest were $^{102}$Ru($n,\gamma$)$^{103}$Ru and $^{102}$Pd($n,\gamma$)$^{103}$Pd.  In the case of ruthenium,
$^{102}$Ru has the highest natural abundance, 32\%, and neutron capture on the other stable isotopes yields either another stable isotope or one with a half-life
substantially shorter than that of $^{103}$Ru.  For palladium, $^{102}$Pd has only 1\% abundance but a relatively large capture cross-section; furthermore, neutron
capture on the other isotopes leads to products with half-lives that ensure they do not compete substantially with $^{103}$Pd decay.  The preparation of the
sources is described in the following sections.   

\subsubsection{\label{RuO2} Natural ruthenium oxide}

We prepared $^{nat}$RuO$_2$ targets by dissolving a sample of 4.5 mg of RuCl$_3\cdot$xH$_2$O powder (99.98\% trace metal basis from Sigma Aldrich, USA) in 185 $\mu$L
of 0.1 M HNO$_3$ and evaporating to dryness under Ar gas. This step converted the ruthenium chloride into ruthenium nitrate. Each sample was then reconstituted
with 5 $\mu$L of 0.1 M HNO$_3$ and 12 mL of anhydrous isopropanol. This solution was then transferred to an electrodeposition cell \cite{Ma13}, and the ruthenium nitrate
was electrochemically deposited using the molecular plating technique \cite{Pa62,Pa64} onto a 25-$\mu$m-thick Al foil backing (99.99\% pure Al from Goodfellow, USA). 
The deposition voltage ranged from 150 to 500 V while the current density was kept between 2 and 7 mA/cm$^2$.  Deposition times ranged from 4 to 5 hours.

After deposition, the targets were baked in atmosphere at 200\,$^{\circ}$C for 30 min to convert the ruthenium nitrate to ruthenium oxide. The resulting targets had
thicknesses between 465 and 545 $\mu$g/cm$^2$ as measured by mass. The plating efficiencies were between 40 and 55\%.  The $^{nat}$RuO$_2$ targets were characterized
with scanning electron microscopy (SEM) to ensure uniformity. An energy-dispersive X-ray spectrometry (EDS) analysis was also performed to verify the elemental
composition, and the EDS spectra showed that Ru and O were indeed the two main components of the target layer. Although the 1:2 ratio expected for Ru:O could not
be verified directly by EDS, this is the most commonly formed oxide of ruthenium. The targets were black in color, as expected of the RuO$_2$ compound.

One of these prepared samples was exposed for 20 hours to a thermal neutron flux of $\sim7.5\times10^{12}\,n$/(cm$^2$\,s) at the TRIGA reactor in the Texas A\&M Nuclear
Science Center.  After removal from the reactor, the sample was conveyed to our measurement location, where counting began a month later, once the shorter-lived 
impurities had decayed away. To our surprise, we discovered that $^{153}$Gd was a prominent impurity, likely because the electrodeposition cell had previously been
used to deposit gadolinium for another experiment.  With a 239-d half-life and x rays in the region of 40 keV, this contaminant proved to be rather troublesome. 

\subsubsection{\label{RuCu} Natural ruthenium/copper foil}

Faced with added experimental uncertainty caused by the gadolinium impurity in our electroplated source, we decided to repeat the measurement using a different target,
one we obtained from the University of Jyv\"{a}skyl\"{a}.  It consisted of 1.1 mg/cm$^2$ of $^{nat}$Ru deposited on 1.3 mg/cm$^2$ of $^{nat}$Cu.  The area of the
ruthenium was about 0.6 cm$^2$; the copper area was about double that.  We activated this target under the same conditions at the TRIGA reactor but for a total of 32
hours.  Once again, we began data acquisition approximately one month later.

\subsubsection{\label{RuCu} Natural palladium foil}

We purchased $^{nat}$Pd metal foil (99.95\% pure from Goodfellow, USA), which was 4-$\mu$m-thick, or 4.8 mg/cm$^2$ in areal density.  The foil was activated for 4.5 hours at
the TRIGA reactor, with data acquisition beginning after more than 2 months.

\begin{figure*}[t]
\epsfig{file=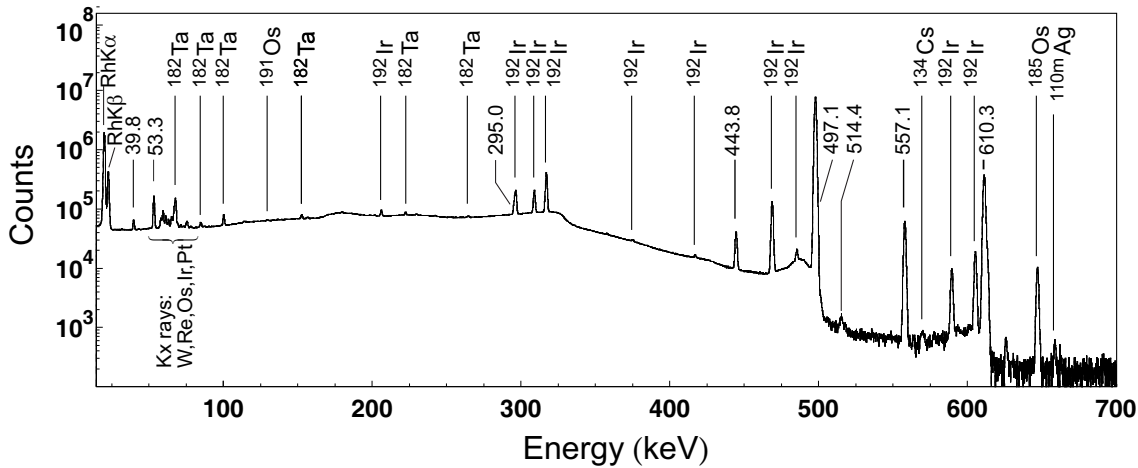,width=15cm}
\caption{Portion of the background-subtracted x- and $\gamma$-ray energy spectrum recorded over a period of 10 days, four months after activation of the Ru/Cu target. 
This was Run 4.  Gamma-ray peaks from the decay of $^{103}$Ru are labeled by their energy; those from impurities are labeled by their $\beta$-decay parent.}
\label{fig2}
\end{figure*}

\subsection{\label{decaymeas} Radioactive decay measurements}

We acquired spectra with our precisely calibrated HPGe detector and with the same electronics used in its calibration \cite{He03}.  Our analog-to-digital converter
was an Ortec TRUMP$^{TM}$-8k card controlled by MAESTRO$^{TM}$ software.  We acquired 8k-channel spectra at a source-to-detector distance of 151~mm, the distance
at which our calibration is well established.  Each spectrum typically covered the energy interval 10-1200 keV with a dispersion of about 0.15 keV/channel.
 
After energy-calibrating our system with a $^{152}$Eu source, we recorded sequential $\sim$12-hour decay spectra, later added together, from each source.  In the case
of the RuO$_2$ target we recorded such spectra, collected into 3 separate ``runs", for a total of 8 days.  For the Ru/Cu source, we recorded spectra in 4 runs, dispersed
over a 3-month period; the total recording time was 33 days.  For Pd, we recorded sequential spectra in 2 runs over a period of 2 months, totaling 21 days.

Because the energy of the rhodium $K_{\alpha}$ x-ray peak is at 20.2 keV and the $\gamma$-ray peak of interest is at 39.8 keV, random pile-up of $K_{\alpha}$ x rays could
seriously interfere with our measurement of the $\gamma$-ray intensity. To remove this possibility, we kept the x-ray counting rate low: for most runs it was less than
50 counts/s, and for no run was it higher than 120 counts/s.  This ensured that the pile-up intensity was well under 1\% of the 39.8-keV peak intensity.

Interspersed among the decay measurements, we recorded sequential room-background spectra for a comparable total time. We could then sum the spectra recorded for each
run and sum the corresponding background spectra, normalize the latter to the same live time as the former and subtract one from the other. The resultant
background-subtracted spectrum from each run was then analyzed, initialy to identify impurities and then, if practical, to extract peak areas for the decay of interest.  
Obviously, changes in the spectrum constituents with time helped us to identify impurities and also to decide which spectra offered the best conditions for clean peak-area
determinations.

We made one further auxiliary measurement.  In the $^{103}$Ru $\beta$-decay measurements, fluorescence in the ruthenium source material interferes with the rhodium x rays
of interest.  Since ruthenium and rhodium x rays cannot be resolved from one another in our HPGe detector, we also measured the Ru/Cu source -- and corresponding background
-- with a 6-mm-diameter, 5.5-mm-deep Si(Li) detector.  The two x-ray groups were cleanly separated in this detector so their relative intensity could easily be established.

\section{\label{sec:analysis} $^{103}$Ru $\beta^-$-decay Analysis}
\subsection{\label{subsec:peakfit} Peak fitting}

A portion of the background-subtracted spectrum from the fourth run recorded with the activated Ru/Cu target is presented in Fig.\,\ref{fig2}: It includes the x- and
$\gamma$-ray peaks of interest from the decay of $^{103}$Ru, as well as a number of peaks from contaminant activities.  

In our analysis of the data, we followed the same methodology as we did with previous source measurements \cite{Ni04,Ni05,Ni07,Ni08,Ni09,Ni14,Ha14,Ni16,Ni17a,Ni17b}.
We first extracted areas for essentially all the  x- and $\gamma$-ray peaks in the background-subtracted spectrum.  Our procedure was to determine
the areas with GF3, the least-squares peak-fitting program in the RADWARE series \cite{Rapc}.  In doing so, we used the same fitting procedures as
were used in the original detector-efficiency calibration \cite{Ha02,He03,He04}.

\subsection{\label{subsec:imp} Impurities}

Once the peak areas (and energies) had been established, we could identify all impurities in each spectrum and carefully check to see if any were known to
produce x or $\gamma$ rays that might interfere with the rhodium $K$ x rays or the 39.8-keV $\gamma$-ray peak of interest.  As is evident from Fig.\,\ref{fig2},
even the weakest peaks were identified. 

In all, for the Ru/Cu source we found 3 weak activities that make a very minor contribution to the rhodium x-ray region.  These are listed in Table\,\ref{table1},
where the contributions are given as percentages of the total number of rhodium x rays recorded, both for Run 1, which was started one month after activation, and for
Run 4, which was started three months later.  The $^{97}$Ru and $^{96}$Tc activities have few-day half-lives and have completely disappeared even by Run 2, while $^{97m}$Tc
is only present as a daughter product of $^{97}$Ru and has a 91-day half-life.  It is relatively stronger in Run 4 than in Run 1.  No impurities interfere
in any way with the $\gamma$-ray peak.

\begin{table}[b]
\caption{\label{table1} The contributions of identified impurities to the energy region of the rhodium $K$ x-ray peaks for Runs 1 and 4 with the Ru/Cu source.  The
contributions are expressed as a percentage of the total number of rhodium x rays.}
\vspace{2mm}
\begin{ruledtabular}
\begin{tabular}{llcc}
 & & \multicolumn{2}{c}{Contribution (\%)} \\
\cline{3-4}\\[-3mm] 
Source & Contaminant &  Run 1 &  Run 4  \\[1mm]
\hline \\[-2mm]
& & &  \\[-4mm]
~~~$^{97}$Ru & Tc $K$ x rays & 0.184(6) & 0 \\
~~~$^{97m}$Tc & Tc $K$ x rays & 0.007(1) & 0.019(2) \\
~~~$^{96}$Tc & Mo $K$ x rays & 0.010(1) & 0  \\
\end{tabular}
\end{ruledtabular}
\end{table}

\begin{figure}[t]
\epsfig{file=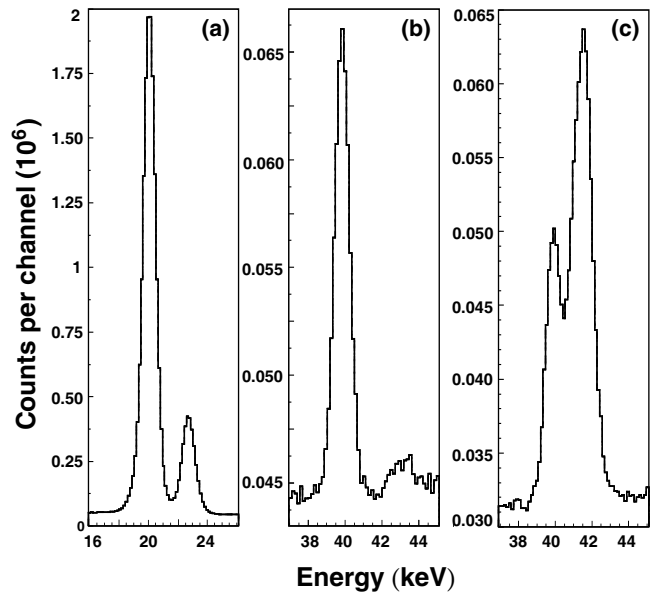,width=8.5cm}
\caption{Spectra for the two energy regions of interest in this measurement. Parts a and b show the rhodium $K$ x rays and the $\gamma$-ray peak at 39.8 keV, both
taken from the full spectrum presented in Fig.\,\ref{fig2} for the Ru/Cu source.  Part c shows the 39.8-keV $\gamma$-ray peak together with europium $K$ x rays from
$^{153}$Gd impurity in our RuO$_2$ source.}
\label{fig3}
\end{figure}

Figure \ref{fig3}, parts a and b, show expanded versions of the two energy regions of interest from the spectrum in Fig.\,\ref{fig2}, one including the rhodium $K$ x
rays and the other, the 39.8-keV $\gamma$ ray.  In both cases, the peaks lie cleanly on a flat background although there is a broad weak peak centered at 42.8 keV, 
which is not far from the $\gamma$-ray peak. This certainly is $not$ random pile-up of two $K$ x rays: it is too high in energy to be two $K_{\alpha}$'s and it shows
no sign of the rate-dependence from run to run that would characterize $K_{\alpha}$-$K_{\beta}$ pile-up.  Instead, it is a mixture of three peaks, one at 42.7 keV
from the decay of $^{182}$Ta, a known impurity (see Fig.\,\ref{fig2}); another from an established transition at 42.6 keV, produced in the decay of $^{103}$Ru; and
a third, at 43.4 keV, which is the Ge-escape peak corresponding to the 53.3-keV transition, also from $^{103}$Ru decay.  The known intensities of these three
transitions fully account for the total counts in this small composite peak.

As an illustrative example of our method for determining $\alpha_K$, the contributing data and corrections are presented for Run 4 in Table \ref{table4}, which
appears later in the text.  The count totals for the $K$ x-ray peaks and for the $\gamma$-ray peak at 39.8 keV appear in the table.  The impurity total of the x-ray
peaks appears below their count total; it corresponds to the percentage breakdown given in Table \ref{table1}

Measurements with the RuO$_2$ source were all taken in an 8-day period beginning one month after activation.  Thus the impurity contributions to the x-ray region of
its decay spectrum were similar to those recorded in Table\,\ref{table1} for Run 1 with the Ru/Cu source, which was accumulated at a similar time after its activation.  
However, as already noted in Sec.\,\ref{RuO2}, the overall spectrum observed with the RuO$_2$ source revealed the presence of a significant gadolinium impurity.  Its
most serious impact was the appearance of europium $K_{\alpha}$ x rays at about 41 keV, originating from the decay of 240-day $^{153}$Gd. The effect can be seen in
part c of Fig.\,\ref{fig3}.  We carefully fitted the spectrum, using a Gaussian function for the 39.8-keV $\gamma$-ray peak and Voigt functions for the $K_{\alpha1}$
and $K_{\alpha2}$ peaks, but inevitably the uncertainty attached to the number of counts in the $\gamma$-ray peak was larger than it was for the Ru/Cu source.

\subsection{\label{subsec:eff} Efficiency ratios} 

As in our previous studies of this type, when we compare the intensities of $K$ x-rays with higher energy $\gamma$ rays, we do not deal separately
with the $K_{\alpha}$ and $K_{\beta}$ x rays.  Scattering effects are quite pronounced at these x-ray energies and they are difficult to account for with an HPGe
detector when peaks are close together, so we use only the sum of the $K_{\alpha}$ and $K_{\beta}$ x-ray peaks.  For calibration purposes, we consider the sum to
be located at the intensity-weighted average energy of the component peaks\footnote[1]{To establish the weighting, we used the intensities of the individual x-ray
components from Table 7a in Ref. \cite{Fi96}.}---20.576 keV for rhodium.

In order to determine $\alpha_K$ for the 39.8-keV $E$3 transition in $^{103}$Rh, we require the efficiency ratio, $\epsilon_{\gamma 39.8}/\epsilon_{K20.6}$, which appears
in Eq.\,(\ref{alpha}).  Following the same procedure as the one we used in analyzing the decay of $^{119m}$Sn \cite{Ni14}, we employ as low-energy calibration the
well-known decay of $^{109}$Cd, which emits 88.0-keV $\gamma$ rays and silver $K$ x rays at a weighted average energy of 22.57 keV.  The latter is close in energy
to the x rays observed in the current measurement.  

To obtain the required ratio we apply the following relation:
\begin{equation}
\frac{\epsilon_{\gamma 39.8}}{\epsilon_{K20.6}} = \frac{\epsilon_{\gamma 88.0}}{\epsilon_{K22.6}} \cdot
\frac{\epsilon_{K22.6}}{\epsilon_{K20.6}} \cdot \frac{\epsilon_{\gamma 39.8}}{\epsilon_{\gamma 88.0}}.
\label{effratio1}
\end{equation} 
We take the $^{109}$Cd ratio $\epsilon_{\gamma 88.0}/\epsilon_{K22.6}$ from our previously reported measurement \cite{Ni14}.  The ratio $\epsilon_{\gamma 39.8}/\epsilon_{\gamma
88.0}$ is close to unity and determined with good precision from our known detector efficiency curve calculated with the CYLTRAN code \cite{He03}, while
$\epsilon_{K22.6}/\epsilon_{K20.6}$ comes from a CYLTRAN calculation as well but requires a short extrapolation beyond the region we have previously calibrated.  The calculated
efficiency drops by less than 4\% from 22.6 to 20.6 keV but to be safe we assign a conservative $\pm$1\% uncertainty.  The values of all four efficiency
ratios from Eq.\,(\ref{effratio1}) appear in the third block of Table \ref{table4}.

\begin{table}[t]
\caption{\label{table2} Our results for the relative intensities of the $\gamma$ rays observed following the $\beta$-decay of $^{103}$Ru, compared with previous
measurements.}
\vspace{2mm}
\begin{ruledtabular}
\begin{tabular}{lllll}
 & \multicolumn{4}{c}{Relative $\gamma$-ray intensities, I$_{\gamma}$} \\
\cline{2-5}\\[-3mm]
 E$_{\gamma}$(keV) & Ref.\,\cite{Ma76} &  Ref.\,\cite{Ch88} &  Ref.\,\cite{Kr10} & This work  \\[1mm]
\hline \\[-2mm]
& & & & \\[-4mm]
39.8  & 0.079(2) & 0.098(9) &  & 0.0752(12) \\
53.3  & 0.42(2) & 0.49(1) & 0.50(12) & 0.420(4)  \\
295.0  & 0.280(9) & 0.333(5) & 0.317(3) & 0.309(4)  \\
443.8  & 0.36(1) & 0.379(4) & 0.373(4) & 0.369(2) \\
497.1  & 100(3) & 100(1) & 100(1) & 100.0(3) \\ 
557.1  & 0.93(3) & 0.95(1) & 0.924(9) & 0.924(3)  \\
610.3  & 6.3(2) & 6.33(5) & 6.15(6) & 6.27(2)  \\
\end{tabular}
\end{ruledtabular}
\end{table}

\begin{table}[b]
\caption{\label{table3} Calculated numbers of x rays generated by the non-isomeric transitions in $^{103}$Rh following the $\beta$ decay of $^{103}$Ru.
Each is expressed as ratio to 1000 counts measured in the 497.1-keV $\gamma$-ray peak.}
\vspace{2mm}
\begin{ruledtabular}
\begin{tabular}{lcccc}
 ~~E$_{\gamma}$ &  Multi- & Mixing &  $\alpha_K$ & N$_K$/N$_{\gamma 497.1}$ \\[1mm]
 (keV) &  polarity & Ratio  &  & ($\times 10^{3}$) \\[1mm]
\hline \\[-2mm]
& & & & \\[-4mm]
53.282(7)  &  $M$1  & & 1.81(1) & 14.1(3)  \\
295.964(10)  &  $M$1+$E$2  & -0.17(1) & 0.0167(1) & 0.095(2)  \\
443.80(2)  &  $E$2  & & 0.00699(1) &  0.048(1) \\
497.083(6)  &  $M$1+$E$2  & -0.368(11) & 0.00458(1) & 8.48(13)  \\ 
557.039(20)  &  $E$2  & & 0.00361(1) & 0.062(1)  \\
610.33(20)  &  $M$1+$E$2  & 0.09(14) & 0.00279(1) & 0.324(5)  \\
\end{tabular}
\end{ruledtabular}
\end{table}

\subsection{\label{subsec:cont} Contributions from other transitions in $^{103}$Rh}

In addition to the isomeric 39.8-keV transition in $^{103}$Rh, there are 6 prompt electromagnetic transitions of appreciable intensity that follow the $\beta$-decay of
$^{103}$Ru, as illustrated in Fig.\,\ref{fig1}. All of them convert to some extent in the $K$ shell so their contributions to the rhodium $K$ x-ray peaks must be accounted for.
To determine their fractional contribution, we need the relative intensities of their $\gamma$ rays and their individual conversion coefficients.  We can, of course, determine
the former from our spectrum (for example, see Fig.\,\ref{fig2}) by making use of the well-established efficiencies of our HPGe detector \cite{Ha02,He03,He04}.

The relative $\gamma$-ray intensities we measure from $^{103}$Ru $\beta$ decay, corrected for coincidence summing, are given in Table\,\ref{table2}, where they are compared
with previous measurements. It
can be seen that our results are the most precise, and agree well with the measurements by Macias {\it et al.}\,\cite{Ma76} and by Krane \cite{Kr10}, but not with the
results of Chand {\it et al.}\,\cite{Ch88}, particularly for the two lowest-energy peaks. Given this situation, we choose to use only our own results rather than an
average over world data.

The known multipolarities and mixing ratios \cite{Fr09} for the 6 prompt transitions are given in Table\,\ref{table3} together with their calculated $K$-conversion coefficients.  The uncertainties
assigned the latter encompass any spread between the two classes of calculation: those that include the atomic vacancy and those that do not.  Combining the result for
$\epsilon_{\gamma 39.8}/\epsilon_{K20.6}$ from Eq.\,\ref{effratio1} with the relative intensity results, I$_{\gamma}$, in Table\,\ref{table2}, we can calculate the x-ray
intensities for all contributing transitions in $^{103}$Rh.  Each result is expressed in Table\,\ref{table3} as a ratio to the number of $\gamma$ rays recorded for the 497.1-keV
transition.  The total contribution from these transitions as determined for Run 4 appears in the first block of Table \ref{table4}.  It constitutes a little over 10\% of the counts
in the rhodium $K$ x-ray peak.

\subsection{\label{subsec:fluor} Ruthenium fluorescence}

\begin{figure}[t]
\epsfig{file=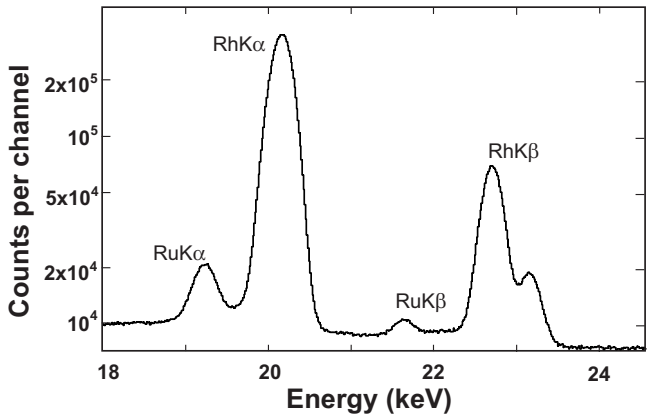,width=8.5cm}
\caption{Spectrum obtained from the Ru/Cu source as measured with a Si(Li) detector. It shows the rhodium $K$ x rays from the decay of $^{103}$Ru as well as the ruthenium $K$ x rays
that result from fluorescence of the source material.}
\label{fig4}
\end{figure}

Though the transition of interest is in rhodium, the source material is predominantly ruthenium.  Because ruthenium has the lower $Z$, rhodium x rays can cause fluorescence
in the source material, creating ruthenium x rays, which only differ in energy by less than 1 keV and consequently cannot be resolved in the HPGe-detector spectrum. To
determine the contribution from fluorescence we recorded the high-resolution spectrum shown in Fig.\,\ref{fig4}, which was accumulated over a period of almost 5 days with the Si(Li)
detector described in Sec.\,\ref{decaymeas}. 

The efficiency of the Si(Li) detector has been thoroughly calibrated \cite{Wa99}; it decreases with increasing energy at 2\%/keV over the energy region covered by the ruthenium and
rhodium $K$ x rays. We also know that the efficiency of our HPGe detector $increases$ with increasing energy at 2\%/keV over the same energy range.  Based on the relative peak areas
in Fig.\,\ref{fig4}, and correcting for the small efficiency differences, we determine that the $K$ x rays of ruthenium constitute 2.92(5)\% of the total intensity of the $K$ x-ray
peaks in the HPGe spectrum.  The correction for Run 4 appears immediately below the total x-ray counts in the first block of Table \ref{table4}. 

\begin{table}[b]
\caption{\label{table4}Corrections to the $^{103}$Ru $K$ x rays and the 39.8-keV $\gamma$ ray, as well as the additional information required to extract a
value for $\alpha_K$.  The data are from Run 4 of the Ru/Cu source measurement and are intended to illustrate the method applied to all runs with both ruthenium sources. }
\vspace{2mm}
\begin{ruledtabular}
\begin{tabular}{lll}
Quantity   &  Value  & Source  \\
\hline \\[-2mm]
\multicolumn{3}{l}{Rh ($K_{\alpha} + K_{\beta}$) x rays}  \\
~~ Total counts & 1.7569(6)$\times 10^7$ & Sec.~\ref{subsec:peakfit}  \\
~~ Ru fluorescence  & -5.13(9)$\times 10^5$  & Sec.~\ref{subsec:fluor}  \\
~~ Impurities  & -3.3(3)$\times 10^3$  & Sec.~\ref{subsec:imp}  \\
~~ Other $^{103}$Rh transitions & -1.806(26)$\times 10^6$   &  Sec.~\ref{subsec:cont} \\
~~ Lorentzian correction & +0.12(2)\%  &  Sec.~\ref{subsec:Lor} \\
~~ Net corrected counts, $N_{K20.6}$  & 1.5264(28)$\times 10^7$  &  \\
\hline \\[-2mm]
\multicolumn{3}{l}{39.8-keV $\gamma$ ray}  \\
~~ Total counts, $N_{\gamma 39.8}$ & 1.505(21)$\times 10^5$ & Sec.~\ref{subsec:peakfit}  \\
\hline \\[-2mm]
\multicolumn{3}{l}{Efficiency ratios (including source attenuation)}  \\ 
~~ $\epsilon_{\gamma 88.0}$/$\epsilon_{K22.6}$ & 1.069(8) & \cite{Ni14} \\
~~ $\epsilon_{K22.6}$/$\epsilon_{K20.6}$ & 1.038(10) & \cite{He03} \\
~~ $\epsilon_{\gamma 39.8}$/$\epsilon_{\gamma 88.0}$ & 1.008(10) & \cite{He03} \\
~~ $\epsilon_{\gamma 39.8}$/$\epsilon_{K20.6}$ & 1.118(18) & \\
\hline \\[-2mm]
\multicolumn{3}{l}{Evaluation of $\alpha_K$}  \\
~~ $N_{K20.6}/N_{\gamma\,39.8}$  &  101.4(14)  & This table  \\
~~ Relative attenuation  &  +0.4(3)\% & Sec.~\ref{subsec:att} \\
~~ $\omega_K$  &  0.809(4)  &  \cite{Sc96}  \\
~~ $\alpha_K$ for 39.8-keV transition  & 140.7(31)   &  Eq.\,(\ref{alpha}) \\
\vspace{-10.pt}
\end{tabular}
\end{ruledtabular}
\end{table} 

\subsection{\label{subsec:Lor} Lorentzian correction}

As explained in our previous papers (see, for example, Ref.\,\cite{Ni04}) we use a special modification of the GF3 program
that allows us to sum the total counts above background within selected energy limits.  To account for possible missed counts outside those limits, the
program adds an extrapolated Gaussian tail.  This extrapolated tail does not do full justice to x-ray peaks, whose Lorentzian shapes reflect the finite
widths of the atomic levels responsible for them.  To correct for this effect we compute simulated spectra using realistic Voigt functions to generate
the x-ray peaks, and we then analyze them with GF3, following exactly the same fitting procedure as is used for the real data, to ascertain how
much was missed by this approach \cite{Ni04}.  The resultant correction factor appears as a percent in the first block of Table\,\ref{table4}.

\subsection{\label{subsec:att} Attenuation in the sample}

Since we are interested in extracting the relative intensities, $N_{K20.6}/N_{\gamma 39.8}$, we need to account for self-attenuation in the source material,
which is slightly different for the 20.6-keV x rays than it is for the 39.8-keV $\gamma$ ray.  The two sources were different as well, one being
$\sim$500-$\mu$g/cm$^2$ RuO$_2$ and the other 1.1-$\mu$g/cm$^2$ ruthenium metal.   Taking the relevant attenuation coefficients from standard tables
\cite{Ch05}, we determined that the x rays suffered 0.13(7)\% more attenuation than the $\gamma$ rays for the RuO$_2$ source, and 0.4(3)\% for the
ruthenium metal source.  The latter value appears in the fourth block of Table\,\ref{table4}

\section{\label{sec:results} $^{103}$Ru $\beta^-$-decay Results}

\subsection{\label{subsec:alphak} $\alpha_K$ for the 39.8-keV transition}

The fourth block of Table\,\ref{table4} contains all the information necessary to evaluate $\alpha_K$ for the 39.8-keV transition from Eq.\,(\ref{alpha}).  Like
everything else in the table, the result appearing on the bottom line is the one obtained from Run 4 with the Ru/Cu source.  The purely statistical contribution
to the total uncertainty on $\alpha_K$ is 2.0, while the systematic contribution -- principally from the efficiency ratio, $\omega_K$ and the attenuation
correction -- is 2.3. Added together in quadrature they become the 3.0 uncertainty value in the table.

\begin{figure}[t]
\epsfig{file=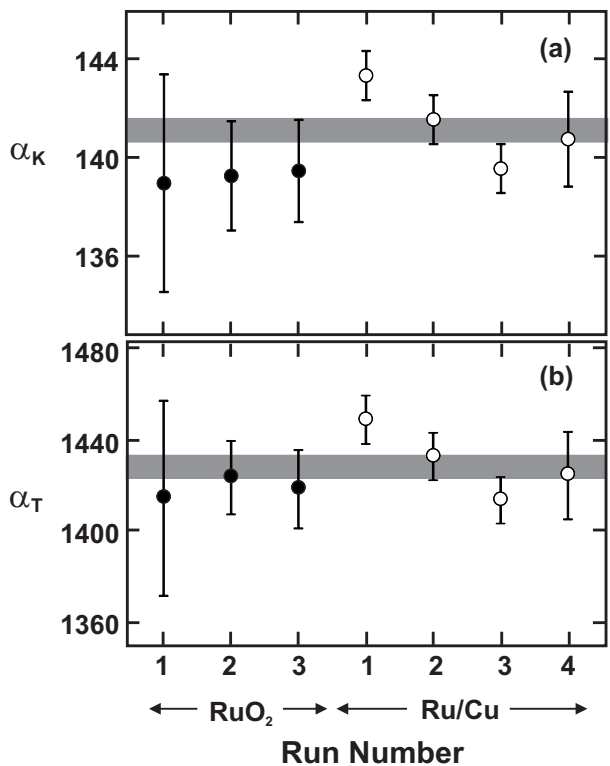,width=8.0cm}
\caption{Plots of the $\alpha_K$ and $\alpha_T$ results from the 7 runs that measured $^{103}$Ru decay.  Only statistical uncertainties are shown.  The
grey bands represent the averages.}
\label{fig5}
\end{figure}

As outlined in Sec.\,\ref{decaymeas}, we took data in 3 runs with the RuO$_2$ source and 4 runs with the Ru/Cu one.  The results from all seven separate measurements
appear in Fig.\,\ref{fig5} with only their statistical uncertainties.  Their average is $\alpha_K$ = 141.1(5).  Adding systematic uncertainties back in we arrive at
the final result:
\begin{equation}
\alpha_{K39.8} = 141.1(23)
\label{alphaK}
\end{equation}
where the uncertainty is dominated by contributions from the efficiency ratios and $\omega_K$.

\subsection{\label{subsec:alphat} $\alpha_T$ for the 39.8-keV transition}

We can now use Eq.\,\ref{alphatot} to determine $\alpha_T$ for the 39.8-keV transition, using the relative $\gamma$-ray intensities from Table\,\ref{table2}
for the 610.3-, 497.1- and 53.3-keV transitions, which feed the 39.8-keV state, combined with their calculated $\alpha_T$ values, which we calculate to be 0.0032(1), 0.0053(1)
and 2.08(3), respectively.  Taking the relative intensity of the 39.6-keV $\gamma$-ray peak from Run 4 with the Ru/Cu source, we obtain $\alpha_{T39.8}$ = 1425(23),
where the counting-statistics contribution to the uncertainty is 20 and that from systematics is 12.

This result along with the results for the other 6 runs appear in Fig.\,\ref{fig5}.  Taking proper account of the statistical and systematic components of the
uncertainties we obtain the average:   
\begin{equation}
\alpha_{T39.8} = 1428(13),
\label{alphaT}
\end{equation}
with systematic uncertainty -- principally from the detector efficiencies -- dominating the error bar.

\begin{figure*}[t]
\epsfig{file=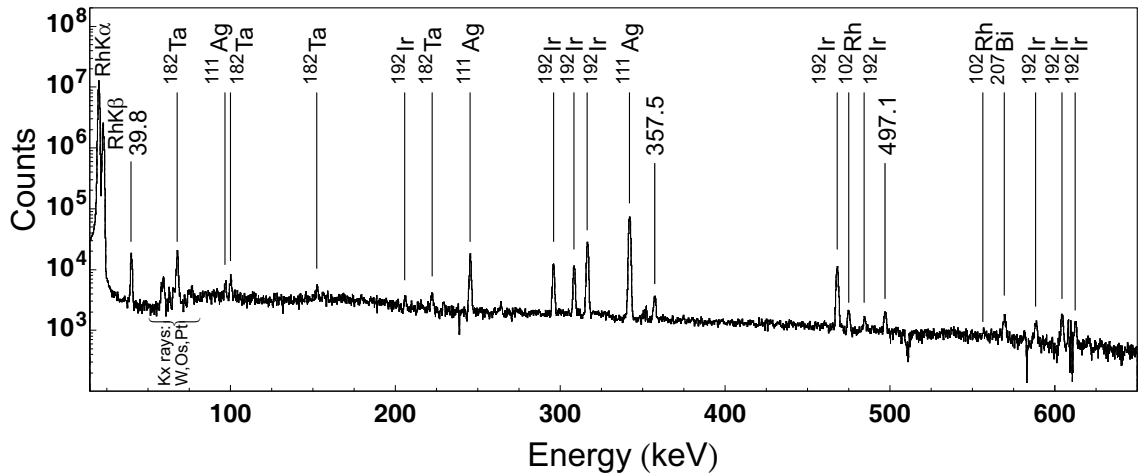,width=15cm}
\caption{Portion of the background-subtracted x- and $\gamma$-ray energy spectrum recorded over a period of 11.6 days, ten weeks after activation of the palladium metal foil. 
Gamma-ray peaks from the decay of $^{103}$Pd are labeled by their energy; those from impurities are labeled by their $\beta$-decay parent.}
\label{fig6}
\end{figure*}

\section{\label{sec:Pdanalysis} $^{103}$Pd $\beta^-$-decay Analysis \& Results}

Numerous x/$\gamma$-ray spectra were recorded, beginning three weeks after the palladium foil had been activated.  It was only much later, though, that the counting rate became
tolerable and the potential for pile-up negligible. Our results are based on two runs: Run 1, which began 10 weeks after activation and continued for 11.6 days; and Run 2, which began after 15 weeks and lasted
8.9 days.

\subsection{\label{subsec:pdimp} Impurities}

A spectrum recorded with the HPGe detector is presented in Fig.\,\ref{fig6}, on which impurities have been identified.  In all, for the palladium source we found two very
weak activities that could contribute to the rhodium x-ray region.  These are listed in Table\,\ref{table5}.  One, $^{102}$Rh, has a half-life of 207 days, which is
considerably longer than 17-day $^{103}$Pd, so its relative contribution to the x-ray region is greater in Run 2 than in Run 1. The second, $^{111}$Ag, is produced as a daughter
of 23-minute $^{111}$Pd and has only a 7.5-day half-life itself.  It has almost vanished by Run 2.   Once again, no impurities interfere in any way with the $\gamma$-ray peak.

\begin{table}[b]
\caption{\label{table5} The contributions of identified impurities to the energy region of the rhodium $K$ x-ray peaks for Runs 1 and 2 with the palladium source.  The
contributions are expressed as a percentage of the total number of rhodium x rays.}
\vspace{2mm}
\begin{ruledtabular}
\begin{tabular}{llcc}
 & & \multicolumn{2}{c}{Contribution (\%)} \\
\cline{3-4}\\[-3mm] 
Source & Contaminant &  Run 1 &  Run 2  \\[1mm]
\hline \\[-2mm]
& & &  \\[-4mm]
~~~$^{102}$Rh & Ru $K$ x rays & 0.025(2) & 0.108(11) \\
~~~$^{111}$Ag & Cd $K$ x rays & 0.0300(5) & 0.0038(1) \\
\end{tabular}
\end{ruledtabular}
\end{table}

As we did for $^{103}$Ru decay, we present the data and corrections from one run -- in this case Run 2 -- as an illustrative example of our analysis.  Table\,\ref{table6} shows
the corresponding count totals for the $K$ x-ray peaks and for the 39.8-keV $\gamma$-ray peak. The impurity total for the x-ray peaks is taken from Table\,\ref{table5} and appears
immediately below the count total.

\begin{table}[t]
\caption{\label{table6}Corrections to the $^{103}$Ru $K$ x rays and the 39.8-keV $\gamma$ ray, as well as the additional information required to extract a
value for $\alpha_K$ + (1+$\alpha_T$)$P_{ec,K}$, which appears in Eq.\,(\ref{alpha2}).  The data are from Run 2 of the palladium source measurement and
are intended to illustrate the method applied to all runs with both ruthenium sources. }
\vspace{2mm}
\begin{ruledtabular}
\begin{tabular}{lll}
Quantity   &  Value  & Source  \\
\hline \\[-2mm]
\multicolumn{3}{l}{Rh ($K_{\alpha} + K_{\beta}$) x rays}  \\
~~ Total counts & 1.3764(5) $\times 10^7$ & Sec.~\ref{subsec:pdimp}  \\
~~ Impurities  & -1.54(15)$\times 10^4$  & Sec.~\ref{subsec:pdimp}  \\
~~ Other $^{103}$Rh transitions & -2.2(8)$\times 10^3$ &  Sec.~\ref{subsec:pdother} \\
~~ Lorentzian correction & +0.12(2)\%  &  Sec.~\ref{subsec:Lor} \\
~~ Net corrected counts, $N_{K20.6}$  & 1.3765(6)$\times 10^7$  &  \\
\hline \\[-2mm]
\multicolumn{3}{l}{39.8-keV $\gamma$ ray}  \\
~~ Total counts, $N_{\gamma 39.8}$ & 1.379(25)$\times 10^4$ & Sec.~\ref{subsec:pdimp}  \\

\hline \\[-2mm]
\multicolumn{3}{l}{Evaluation of $\alpha_K$ + (1+$\alpha_T$)$P_{ec,K}$}  \\
\multicolumn{3}{l}{for the 39.8-keV transition}  \\
~~ $\epsilon_{\gamma 39.8}$/$\epsilon_{K20.6}$ & 1.118(18) & Table\,\ref{table4} \\
~~ $N_{K20.6}/N_{\gamma\,39.8}$  &  997.8(18)  & This table  \\
~~ Relative attenuation  &  +0.3(2)\% & Sec.~\ref{subsec:pdother} \\
~~ $\omega_K$  &  0.809(4)  &  Sec.~\ref{overview}  \\
~~ $P_{ec,K}$  &  0.8595(10)  &  \cite{Sc96}  \\
~~ $\alpha_K$ + (1+$\alpha_T$)$P_{ec,K}$  & 1383(34)   &  Eq.\,(\ref{alpha2}) \\
\vspace{-10.pt}
\end{tabular}
\end{ruledtabular}
\end{table}

\subsection{\label{subsec:pdother} Other Corrections}

As is evident from Fig.\,\ref{fig1} the electron-capture decay of $^{103}$Pd leads almost exclusively to the 39.8-keV transition of interest; no competing
transition is stronger than 0.03\% of its intensity \cite{Fr09}.  Even so, two weak $\gamma$ rays from other transitions in $^{103}$Rh are visible in Fig.\,\ref{fig6}
so, for the sake of completeness, we evaluated the contribution of all competing transitions in that nucleus \cite{Fr09} to the rhodium $K$ x rays.  The result, though
essentially negligible, appears in the first block of Table\,\ref{table6}. 

Because the palladium source material has higher $Z$ than rhodium and there are no strong $\gamma$ rays to contend with, fluorescence is not an issue for this decay.

The effect of attenuation on the $^{103}$Pd-decay measurement is similar to that for the $^{103}$Ru source.  Following the same procedure as described in Sec.\,\ref{subsec:att}
we find that the x rays suffered 0.3(2)\% more attenuation than the 39.8-keV $\gamma$ rays.  This result appears in the third block of Table\,\ref{table6}.

\subsection{\label{subsec:pdresult} Result for $\alpha_K$ + (1+$\alpha_T$)$P_{ec,K}$}

With the information in Table\,\ref{table6} and the use of Eq.\,(\ref{alpha2}) we derive the result  for $\alpha_K$ + (1+$\alpha_T$)$P_{ec,K}$, which appears on the bottom
line of the table.  This of course is the result for Run 2 only.  Of the $\pm$34 uncertainty quoted, the purely statistical contribution is 25 and the systematic contribution
-- again principally from the efficiency ratio, $\omega_K$ and the attenuation correction -- is 22.  If we combine the result in Run 1, taking care to keep the statistical and
systematic uncertainty components separate, we obtain the final result:
\begin{equation}
\alpha_K + (1+\alpha_T)P_{ec,K} = 1383(28)
\label{alphaKT}
\end{equation}
where the uncertainty is about equally shared between counting statistics and systematic effects.

\section{\label{sec:disc} Discussion}

We have obtained three experimental results, presented in Eqs.\,(\ref{alphaK}-\ref{alphaKT}), which involve two quantities we seek to determine, $\alpha_{K39.8}$ and
$\alpha_{T39.8}$.  A linear least-squares fit can in principle yield optimum values for these two quantities, which best satisfy all three measurements; however, the interplay of
statistical and systematic uncertainties makes this process problematical. If we make the fit with only statistical uncertainties attached to the measurements
and then add the systematic uncertainties back onto the fitted results, we obtain $\alpha_{K39.8}$ and $\alpha_{T39.8}$ values that do not significantly differ from those
appearing in Eqs.\,(\ref{alphaK}) and (\ref{alphaT}).  We choose therefore to consider the results in those equations as final, and view the result in Eq.\,(\ref{alphaKT}) as
providing independent confirmation, noting that if we substitute the results from Eqs.\,(\ref{alphaK}) and (\ref{alphaT}) into the left side of Eq.\,(\ref{alphaKT}), we
obtain the value 1369(25), which agrees well with the separate measurement on the right side of that equation.

There have been six previous measurements of $\alpha_K$ for the 39.8-keV transition in $^{103}$Rh, of which the most precise yielded 127(6) \cite{Cz75} and 153(6) \cite{Sa99}.
Both disagree with our even more precise result of 141.1(23) but, curiously, their average agrees completely.  The first of these measurements, from 1975, used a NaI(Tl)
detector having 25\% resolution at 20 keV and, with this resolution, their spectrum would give no hint of $K$ x-ray pile-up contributing to their $\gamma$-ray peak, yet there
is no mention of pile-up in their publication and there is no suggestion that they took steps to avoid it.  A low value for $\alpha_K$ is just what one would expect from the
presence of an undiagnosed intruder in the $\gamma$-ray peak. The second measurement, made in 1999, made use of a mini-orange spectrometer for conversion electrons and an
HPGe detector for the 39.8-keV $\gamma$, with the two detector-efficiency functions connected by another transition in $^{103}$Rh with a known ICC.  Except that this requires
a much more complicated and error-prone calibration procedure than ours, there is no evident reason for the measurement to be flawed.

The value for $\alpha_{T39.8}$ has only been measured twice before, with the results 1430(89) \cite{Va79} and 1531(30) \cite{Cz75}. Again, it is Ref.\,\cite{Cz75} that
disagrees with our current measurement, 1428(13). There is no obvious reason for the disagreement but it should be noted that the technique used in the older measurement
was rather complex, while ours in essence depended only on the relative intensities of two peaks in a well-calibrated HPGe spectrum.

Before we compare our results with theory, it is important to establish the energy of the 39.8-keV transition as precisely as possible since the calculated ICCs are
sensitive to the transition energy.  There are three comparably precise and consistent measurements in the literature: One used a curved-crystal spectrometer to
obtain 39.755(12) keV \cite{Ra69}; the others used electron spectrometers to extract 39.748(8) keV \cite{Gr69} and 39.762(16) keV \cite{Pe70}.  We use their
weighted average, 39.752(6) keV.

In Table\,\ref{table7} our results are compared with three different theoretical calculations under two separate assumptions for the multipolarity mix of the transition. 
All three calculations were made within the Dirac-Fock framework, but one ignores the presence of the $K$-shell vacancy while the other two include it using different
approximations: the frozen-orbital (FO) approximation, in which it is assumed that the atomic orbitals have no time to rearrange after the electron's removal; and the
SCF approximation, in which the final-state continuum wave function is calculated in the self-consistent field (SCF) of the ion, assuming full relaxation of the ion
orbitals.  For a full description of the various models used to determine the conversion coefficients, see Ref.\,\cite{Ni04}.

\begin{table}[b]
\caption{\label{table7}Comparison of the measured $\alpha_K$ and $\alpha_T$ values for the 39.752(6)-keV $E$3 transition in $^{103}$Rh with calculated values based on
three different theoretical models, one that ignores the $K$-shell vacancy and two that deal with it either in the ``frozen-orbital" (FO) approximation or the self-consistent
field (SCF) approximation (see text).  The uncertainties on the calculations reflect the uncertainty in the measured transition energy.  Shown also are the percentage
deviations, $\Delta$, from the experimental value calculated as (experiment-theory)/theory.  Calculated values are given, both for a pure $E$3 transition and for an $E$3+$M$4
transition with a mixing ratio of $\delta$=0.02.}
\vspace{2mm}
\begin{ruledtabular}
\begin{tabular}{lllll}
\multicolumn{1}{l}{Model}  & \multicolumn{1}{c}{~~$\alpha_K$} & \multicolumn{1}{c}{~~$\Delta$(\%)} & \multicolumn{1}{c}{~~$\alpha_T$} & \multicolumn{1}{c}{~~$\Delta$(\%)} \\
\hline \\[-3mm]
Experiment & 141.1(23)  & & 1428(13) &  \\
Theory: & & \\
  &  &  \\[-3mm]
~a)Pure $E$3  &  &  \\
~~~~No vacancy  & 127.5(1) & +10.7(18) & 1388(2) & +2.9(9) \\
~~~~Vacancy, FO  & 135.3(1) & +4.3(17) & 1404(1) & +1.7(9) \\
~~~~Vacancy, SCF  & 133.2(1) & +5.9(17) & 1399(1) & +2.1(9) \\
  &  &  \\[-3mm]
~b)$E$3+$M$4, $\delta$=0.02  &  &  \\
~~~~No vacancy  & 131.3(1) & +7.5(18) & 1410(2) & +1.3(9) \\
~~~~Vacancy, FO  & 139.4(1) & +1.2(17) & 1426(2) & +0.1(9) \\
~~~~Vacancy, SCF  & 137.2(1) & +2.8(17) & 1421(2) & +0.5(9) \\
\vspace{-10.pt}
\end{tabular}
\end{ruledtabular}
\end{table}

Currently, the 39.8-keV transition is taken by the evaluator \cite{Fr09} to be pure $E$3 in multipolarity, presumably based on a 1970 measurement of $L$-subshell
conversion-line intensities \cite{Pe70}, from which the authors deduced that any $M$4 admixture had to be less than 0.04\%. If we accept that the $M$4 admixture in the
transition is exactly zero, then we see from Table\,\ref{table7} that the comparison between experiment and theory for both $\alpha_K$ and $\alpha_T$ strongly disagrees
with the calculation that ignores the atomic vacancy but also disagrees, albeit by a smaller amount, with the calculations that include provision for the vacancy.

The table also shows, though, that if we assume the tiny 0.04\% $M$4 admixture ($\delta$ = 0.02) allowed by the upper limit set in Ref.\,\cite{Pe70}, then agreement
with the theory that includes the vacancy becomes excellent, while disagreement with the no-vacancy approach remains significant.

In either case we can conclude that, once again, experiment rules out ICC calculations that do not take account of the atomic vacancy.  In this respect, our new result is
consistent with our previous eight precise $\alpha_K$ measurements on $E$3 and $M$4 transitions in $^{111}$Cd \cite{Ni16}, $^{119}$Sn \cite{Ni14,Ha14}, $^{125}$Te
\cite{Ni17b}, $^{127}$Te \cite{Ni17a}, $^{134}$Cs \cite{Ni07,Ni08}, $^{137}$Ba \cite{Ni07,Ni08}, $^{193}$Ir \cite{Ni04,Ni05} and $^{197}$Pt \cite{Ni09}, all of which
disagreed -- some, as this case, by many standard deviations -- with the no-vacancy calculations.   

At the same time, our new result for $\alpha_K$ differs from the previous measurements in that it disagrees -- by more than two standard deviations -- with the
vacancy-included calculations as well if the transition is assumed to have unique multipolarity.  However, we have shown that agreement can be restored for both $\alpha_K$ and
$\alpha_T$ if we assume that the 39.8-keV transition contains a very small admixture of $M$4, an amount that is not ruled out by any other known data.

Finally, if we take the position that the need for the vacancy to be included in ICC calculations has already been proven by our previous eight measurements, then we
can use these calculations to determine the mixing ratio that best fits the data for $\alpha_K$ and $\alpha_T$. Doing so, we determine the mixing ratio for the 39.8-keV
transition to be $\delta$ = 0.023(5).

\section{Conclusions}

This measurement was originally undertaken to extend our systematic tests of internal conversion theory to a lower-$Z$ nucleus.  The eight $E$3 and $M$4 transitions we had
studied previously all strongly favored ICC calculations that took account of the atomic subshell vacancy left by the conversion process, and we were seeking to establish
the validity of this conclusion over as wide a region of the nuclear chart as possible.  It might be said that, with the current result, we have only partially succeeded: The
new results for $\alpha_K$ and $\alpha_T$ do in fact disagree with the calculation that ignores the vacancy, but the agreement with the preferred calculation is not entirely
satisfactory either. 

But this is only if the transition is taken to be pure $E$3 in character.  A small $M$4 admixture is allowed within previous experimental limits, and its inclusion simultaneously
brings both $\alpha_K$ and $\alpha_T$ into agreement with the vacancy-included calculations.  This constitutes very strong circumstantial evidence that the calculations are indeed
correct and that the 39.8-keV transition is a mixed $E$3 + $M$4 transition with $\delta$ = 0.023(5).

A scan of NuDat records at the NNDC website \cite{NNDC}, covering the whole nuclear chart, yields only seven known transitions of potentially mixed $E$3 + $M$4 character, none of which has
a measured mixing ratio, $\delta$, with an uncertainty that does not overlap zero.  It appears that the transition we have measured in $^{103}$Rh is the first one ever determined to have
a definitively non-zero value.  Given that this mixing ratio corresponds to a mere 0.05\% admixture, it is perhaps not surprising that previous measurements have not been sensitive
enough to observe such a tiny effect.

It would, of course, be very valuable to have an independent measurement of the mixing ratio for the 39.8-keV transition by a different technique.

\begin{acknowledgments}

We thank the Texas A\&M Nuclear Science Center staff for their help with the neutron activations; and T. Eronen from the University of Jyv\"{a}skyl\"{a} for providing us with
the material used for our Ru/Cu source.  This report is based upon work supported by the U.S. Department of Energy, Office of Science,
Office of Nuclear Physics, under Award Number DE-FG03-93ER40773, and by the Robert A. Welch Foundation under
Grant No.\,A-1397.

\end{acknowledgments}

\end{document}